\documentstyle[aps,twocolumn,floats,epsfig]{revtex}
\begin{document}
\bibliographystyle{prsty}
\wideabs{
\begin{flushleft}
{\small \em submitted to}\\ {\small PHYSICAL REVIEW B \hfill
VOLUME {\normalsize XX}, NUMBER {\normalsize XX} $\qquad\qquad$
\hfill MONTH {\normalsize XX} }
\end{flushleft}

\author{E. del Barco, J.M. Hernandez, and J. Tejada.}
\address{Dept. de F\'\i sica Fonamental, Universitat de Barcelona, Diagonal 647,\\
08028 Barcelona, Spain.}
\author{N. Biskup, R. Achey, I. Rutel, N. Dalal, and J. Brooks}
\address{NHMFL, CM/T group 1800
E.P.Dirac dr. Tallahassee, FLORIDA 32310}
\title{High frequency resonant experiments in Fe$_8$ molecular clusters.}
\maketitle

\date{Received 4 November 1999}

\begin{abstract}
Precise resonant experiments on Fe$_{8}$ magnetic clusters have
been conducted down to 1.2 K at various tranverse magnetic fields,
using a cylindrical resonator cavity with 40 different frequencies
between 37 GHz and 110 GHz. All the observed resonances for both
single crystal and oriented powder, have been fitted by the
eigenstates of the hamiltonian ${\cal H}=-DS_z^2+ES_x^2-g\mu
_B{\bf H}\cdot {\bf S}$. We have identified the resonances
corresponding to the coherent quantum oscillations for different
orientations of spin S = 10.
\end{abstract}
\smallskip
\begin{flushleft}
PACS number(s): 75.45+j
\end{flushleft}
 }
At low temperature magnetic molecules with spin S = 10, Mn$_{12}$
and Fe$_8$, are equivalent to a single domain particle with a
constant magnetic moment, $m= 20 \mu_B$. The orientation of this
magnetic moment freezes, however, along one of two easy directions
when temperature becomes small compared to the energy barrier
height due to magnetic anisotropy. The discovery of resonant spin
tunneling was first achieved by performing both hysteresis
measurements and relaxation experiments on Mn$_{12}$-acetate below
that blocking temperature \cite{Friedman,Hernandez,Thomas,Libro}.
More recently new experimental proofs of this effect have been
obtained by using different experimental techniques
\cite{Fominaya,Luis,Sales,Perenboom}. Comprehensive theories for
the tunneling rate in the presence of phonons were developed which
explain quantitatively the experimental results
\cite{Garanin,Fernandez,Villain}. Very recently two new
experiments have brought more excitement to this field.
Wernsdorfer et al. \cite{Wernsdorfer} have detected the
suppression of the spin tunneling rate in Fe$_8$ due to the
non-Kramers topological quenching of tunneling noticed by Garg
\cite{Garg}. Del Barco et al. \cite{delBarco} have detected the
coherent quantum oscillations of the spin 10 in Fe$_8$ by
performing resonance experiments at 680 MHz. In this paper we
report the results of high frequency experiments on Fe$_8$,
between 37 GHz and 110 GHz, in the Kelvin regime.This experiment
follows the pioneering idea of Awschalom et al. \cite{Awschalom}
on the search of quantum coherence of the spin in nanomagnets by
performing resonant experiments.

The material studied is composed by Fe$_8$ single crystals
synthesized according to ref. \cite{Wiedghardt}. The magnetic
susceptibility data support the non existence of paramagnetic
impurities and the measured crystal cell parameters are in full
agreement with those previously published \cite{Wiedghardt}. The
experiments were carried out on a single crystal of 2 mm length
and on a oriented powder sample composed of small crystallites of
1 $\mu$m average size. The orientation of the the powder was done
by solidifying an epoxy with the Fe$_8$ microcrystallites buried
inside and applying during 12 h, the solidifying time of the
epoxy, a magnetic field of 5.5 T. The nominal composition, [({\it
C}$_6${\it H}$_{15}${\it N}$_3$)$_6${\it Fe}$ _8$($\mu _3$-{\it
O})$_2$($\mu _2$-{\it OH})$_{12}$({\it Br}$_7$({\it H}$_2$ {\it
O})){\it Br}$_8${\it H}$_2${\it O}], was checked by chemical and
infrared analysis. The single crystal was placed on the cavity of
the resonator with its hard plane ($xy$ plane) mostly parallel to
the applied field. It was nevertheless a small misalignment angle,
$\theta$, between the direction of the applied magnetic field and
the hard plane. In fact this angle is the only fitting parameter
we used in our fitting procedure.

The occurrence of resonant spin tunneling in Fe$_8$ was first
detected by Sangregorio et al. \cite{Sangregorio}. The spin
hamiltonian has been  deduced from EPR experiments \cite{Barra}.
The ac susceptibility data published by Zhang et al.
\cite{XiXiang} on an oriented powder sample of Fe$_8$ are in
agreement with those above mentioned. Our materials were
characterized by performing dc and ac magnetic measurements on
both single crystal and oriented powder. Both samples show
periodic steps in the magnetic data, separated by a period of 0.24
T. We have also performed magnetic susceptibility measurements at
different fields down to 100 mK. The zero field cooled (ZFC) and
field cooled (FC) curves merge at 0.8K. A Curie-Weiss behaviour is
observed between this temperature and few K which corresponds to
the only population of the level S = 10. The Curie temperature
deduced from this low temperature susceptibility data is 0.5 K
which is the measure of the interaction between the molecules. At
higher temperature strong deviations from the Curie-Weiss is
observed due to the population of levels with S $<$ 10 as it was
first experimentally demonstrated by Delfs et al. \cite{Delfs}.

To the first approximation, the Hamiltonian of Fe$_8$ is
\cite{Sangregorio,Barra}:

\begin{equation}
\label{Hamiltonian}{\cal H}=-DS_z^2+ES_x^2-g\mu _B{\bf H}\cdot
{\bf S}+C(S_+^4+S_-^4)
\end{equation}
\begin{figure}
\centerline{\epsfig{file=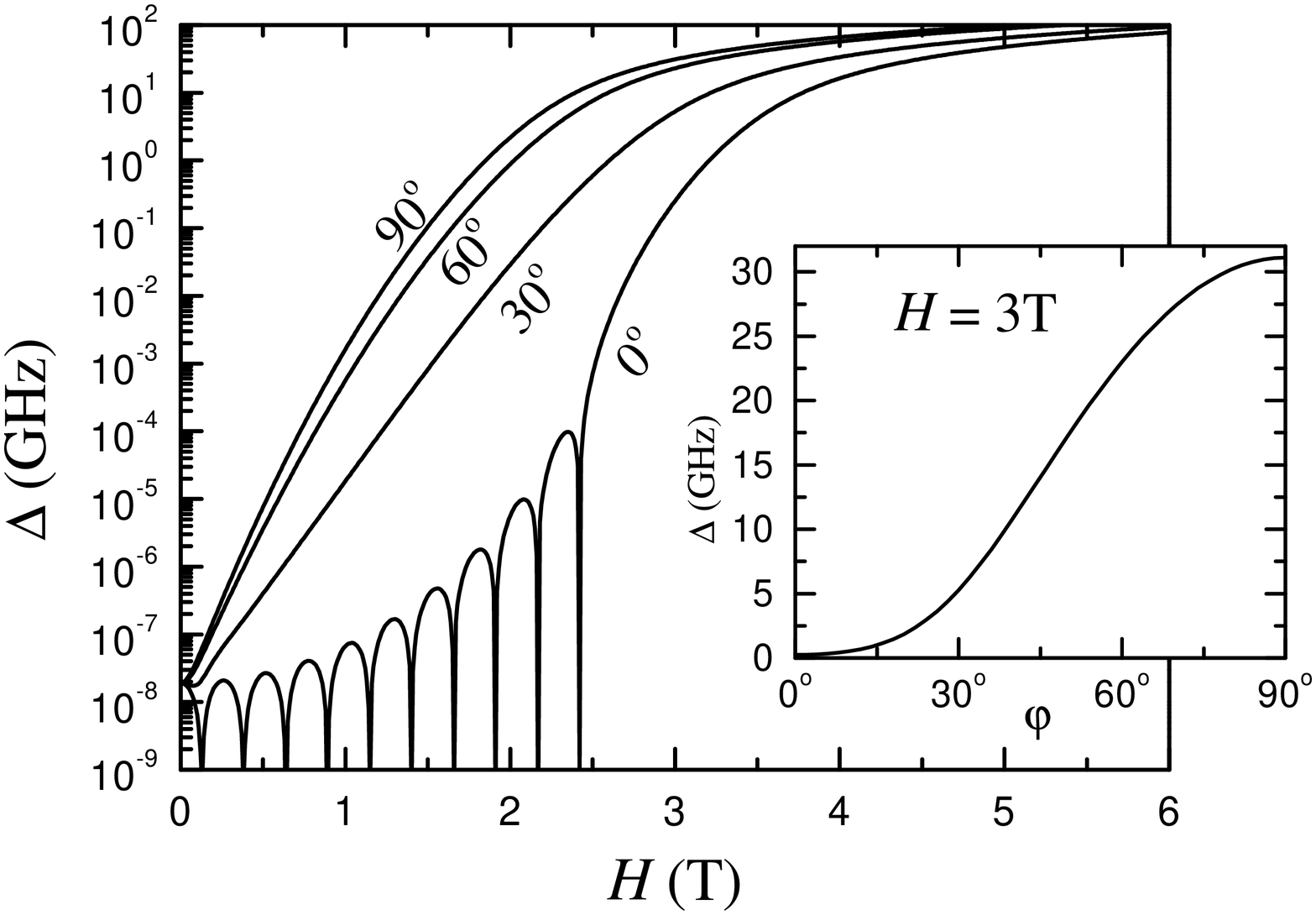,width=8truecm}}
\caption{\small Dependence of the ground-state splitting on the
transverse field in Fe$_8$ for different angles $\varphi$. The
inset shows the angular dependence of the splitting at a fixed
value of the field $H$ = 3 T.}
\end{figure}

where $z$ and $x$ denote the easy and hard axes respectively. The
values of $D =$ 0.229 K, $E =$ 0.093 and $C =$ 2.9 10$^{-5}$ K are
known from EPR, neutron spectroscopy and magnetic relaxation
experiments \cite{Sangregorio,Barra,Caciuffo,Wernsdorfer}. H is
the applied magnetic field. The component of the applied magnetic
field parallel to the easy axis direction, called longitudinal
component, $H_\|$ = $H$ $\sin \theta$, changes the barrier height
between the two classical  spin orientations, while the component
of the field on the hard plane, transverse component, $H_\bot$ =
$H$ $\cos \theta$, affects the overlapping of the respective wave
functions, which determines the quantum splitting of the
degenerate spin states. The quantum splitting, $\Delta$, and
consequently the rate of resonant tunneling between the spin
levels depend on both the magnitude of the transverse component
$H_\bot$ and its angle $\varphi $ with the hard axis. The
dependence of the quantum splitting of the ground state, $S_z =$
10, on the applied magnetic field, $\Delta (H_\bot)$ is shown in
figure 1. The angular dependence of $\Delta$ ($\varphi $) at a
fixed value of the transverse component of the field, $H_\bot =$ 3
T, obtained by numerical diagonalisation of the Hamiltonian (1) is
plotted in the inset of Figure 1. Because of the shape of the
function $\Delta$ ($\varphi $), for a sample with hard axis
oriented at random, that is with not preference for any angle
$\varphi $, there are two values of $\Delta$ for which the density
of states has a peak. These are the values of the splitting
corresponding to $\varphi =$ 0 and $\varphi = \frac \pi 2$.

\begin{figure}
\centerline{\epsfig{file=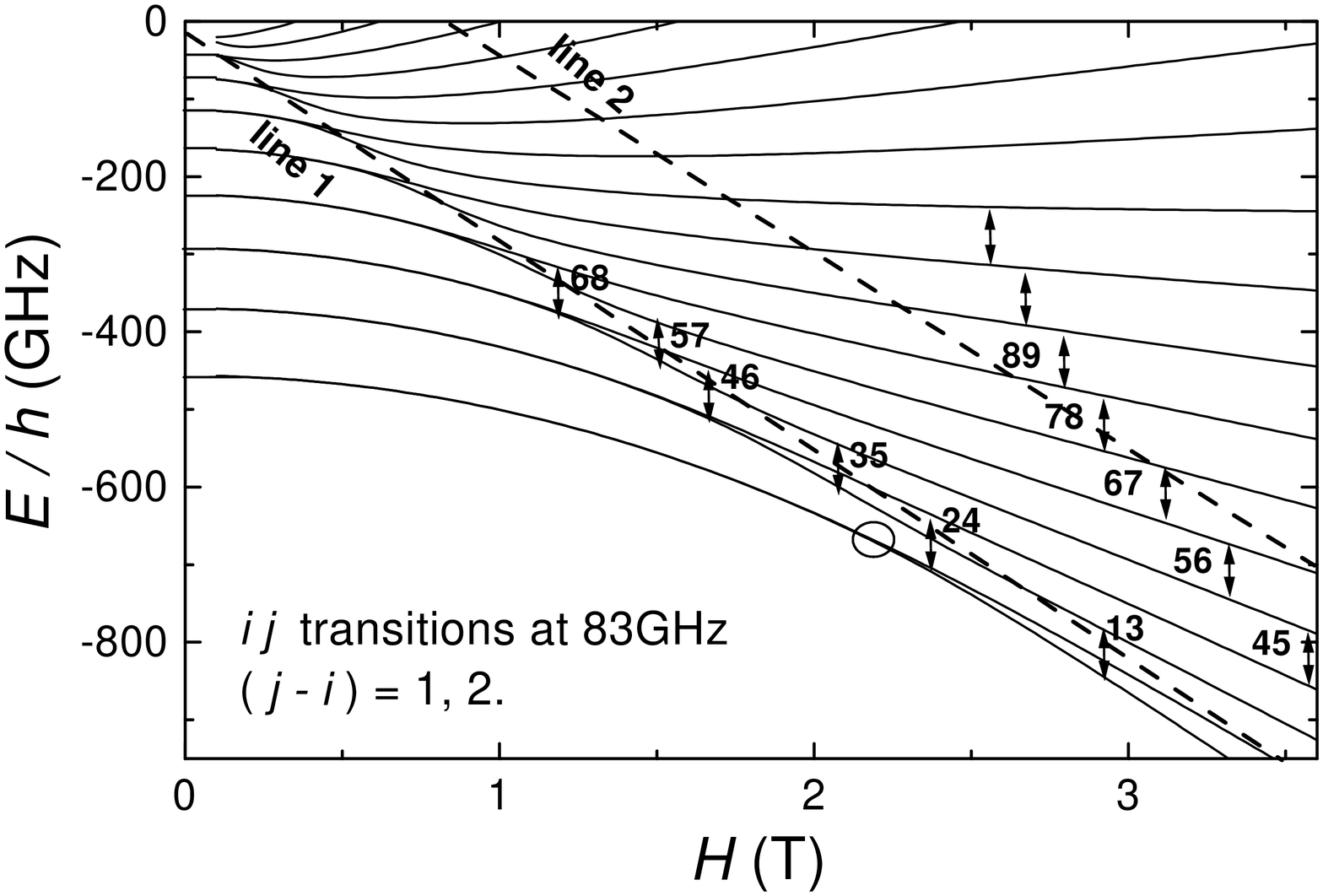,width=8truecm}}
\caption{\small Evolution of the spin levels in the two wells on
the intensity of the transverse field for $\varphi$ = $\frac \pi
2$. Lines 1 and 2 limit the transition zone between the quantum
splittings of the $S_z$ levels (low frequency) and the strongly
mixed $S_z$ states (high frequency). Open circle at 2.25 T in the
lowest level indicates the absorption peak corresponding to a
frequency of 680 MHz extracted from ref. 14 }
\end{figure}

The evolution of the spin levels in the two wells as a function of
the intensity of the transverse component of the field, as deduced
from the diagonalisation of the spin hamiltonian, is shown in
Figure 2. The resultant spin levels on the left side of line 1
correspond to the quantum splittings of the different $S_z$ levels
due to tunneling. In the case of the lowest $\mid S_z\mid  =$ 10
level, the tunneling gives rise to the true ground state and the
first excited state which are separated by the energy
$\Delta_{10}$. The probability of the spin of each Fe$_8$ molecule
to have a certain orientation ${\bf S}$, at a moment of time $t$
is given by $P(t)= \cos(\Gamma t)$. Consequently our system of
Fe$_8$ molecules placed in an $ac$ field of frequency $\omega$
will show a resonance in the power spectrum if $\omega =
\Delta_{10}$/$\hbar = \Gamma$. To observe this resonance and those
associated with different $S_z$ levels, the dissipation should be
small compared to the quantum splitting. Dissipating destroys the
coherence of spin quantum oscillations after a certain decoherence
time. This is the reason why it is necessary to carry out
experiments at high fields (large splittings) and, consequently,
high frequencies. The transverse component of the magnetic field
must be, nevertheless, smaller than the anisotropy field, $H_a$,
to have degenerate spin levels. These are the levels on the left
on line 1 in Figure 2. For $H_\bot$ values greater than the
anisotropy field the labeling of the spin levels is more
complicate as they result from strongly mixing of the $S_z$
states. That is, in this high-field regime, on the right of line
2, the appropriate eigenstates are $S_x$ and $S_y$ depending on
the direction of the applied magnetic field. Between these two
lines, there is an intermediate region in which for a given
$H_\bot$ value there are both types of levels.

In the absence of dissipation, the contribution of each Fe$_8$
crystallite to the imaginary part of the susceptibility is
proportional to $\delta (\omega -\frac {\Delta [\varphi ,H_\bot]}
\hbar)$. However, the total imaginary part of the susceptibility
is,

\begin{equation}
\label{suscept} \chi'' \propto \int_{0}^{\pi} g(\varphi) \delta (
\omega - \frac {\Delta [\varphi ,H_\bot]} \hbar ) d\varphi \;\; ,
\end{equation}

\noindent where $g(\varphi )$ is the distribution of crystallites
on $\varphi $. For a single crystal, having all its molecules with
$x$ axis exactly at the same angle, the amplitude $A$ of the
absorption of electromagnetic radiation must have only one peak
corresponding to $\Delta [\varphi ,H_\bot]=\omega $. For an
oriented powder where there is a random orientation of the $x$
axis of each crystallite, Eq. \ref{suscept} can be rewritten as

\begin{equation}
\label{suscept2}  \chi'' \propto \int_{0}^{\infty} \delta ( \omega
- \frac {\Delta [\varphi ,H_\bot]} \hbar ) \left (\frac {d\Delta}
{d\varphi} \right )^{-1} d\Delta= \left.\left (\frac {d\Delta}
{d\varphi} \right )^{-1}\right |_{\Delta=\hbar \omega} \;\; .
\end{equation}

\noindent Therefore, there are two field values, solutions of the
equations $\Delta [0,H_{\bot 1}]=\omega$ and $\Delta [\frac \pi 2
,H_{\bot 2}]=\omega $, at which the amplitude of the absorption is
maximal.

The high frequency  resonance experiments have been carried out
using the AB millimeter wave vector network analyzer
(MVNA)\cite{Abmm}. The base frequency obtained from this source
(range 8 - 18 GHz) is multiplied by Q, V and W Schottkey's diodes
to obtain frequency range used in our experiment (37 - 109 GHz).
The sample, a single Fe$_8$ crystal or the oriented powder, is
placed on the bottom of the cylindrical resonant cavity, halfway
between its axis and perimeter. The applied dc magnetic field is
parallel to the cavity axis and approximately perpendicular to the
easy (c) axis of the crystal. The experiment frequencies are
TE$_{0np}$ ($n$, $p=$ 1, 2, 3, ...) which are the resonant
frequencies for the cavity used. Resonance Q-factor varies from
20000 at TE$_{011}$ mode (41.6 GHz) to few thousand at higher
frequencies. Due to high resonance sensitivity, we may detect all
absorption peaks suggested theoretically by the diagonalization of
the Hamiltonian of Equation 1.

\begin{figure}
\centerline{\epsfig{file=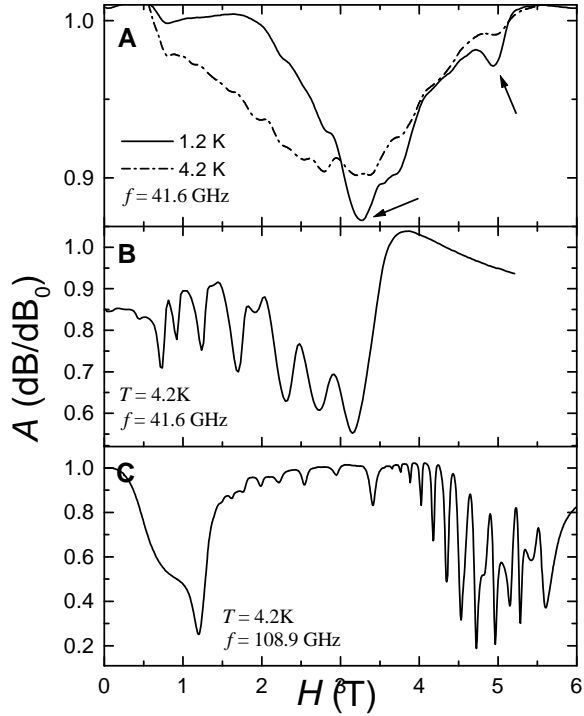,width=8truecm}}
\caption{\small Oriented powder (A) and single crystal absorption
vs. magnetic field spectra (C,D) at various frequencies and
temperatures. Arrows in fig. A point to the two peaks
corresponding to $\varphi = \frac \pi 2$ ($H$ = 3.2 T) and
$\varphi = 0$ ($H$ = 4.9 T) orientations of the field on the hard
plane.}
\end{figure}

The data for the Fe$_8$ single crystal were obtained at $T=$ 4.2 K
while $T$ was varied between 1.2 K to 10 K for the oriented powder
sample. Figure 3 shows the absorption curves for the two samples;
figure 3A shows the spectra for the powder sample at two different
temperatures and $f=$ 41 GHz. The most pronounced peaks shift to
lower fields as the temperature is increased. Figures 3B and 3C
show the resonance spectrum for the single crystal at two
different frequencies; The resonant fields shift to higher fields
when increasing the frequency. The resonance spectrum for the
single crystal at $T=$ 4.2 K and $f=$ 41 GHz, has only peaks for
field values lower than 3.2 T, while the spectrum for the oriented
powder at the same temperature and frequency shows peaks until 5.2
T.

In Figure 4 we collect the values for the resonant peaks for the
single crystal  when using 40 different frequencies (solid
circles). We also show in this figure (open squares) the values of
the resonances for five different frequencies and $T=$ 1.2 K for
the oriented powder. In this case we have selected the two most
pronounced peaks for each frequency.
\begin{figure}
\centerline{\epsfig{file=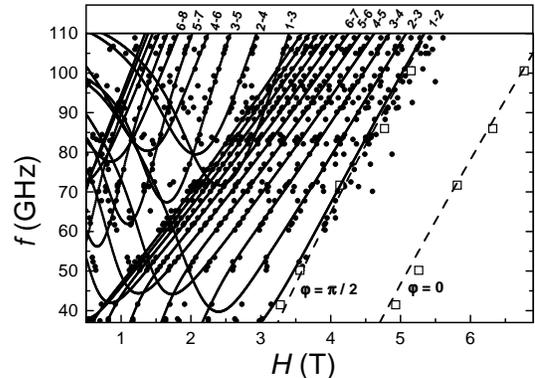,width=8truecm}}
\caption{\small  Absorption resonant peaks for oriented powder
(open squares) and single crystal (solid circles). The lines
represent the theoretical calculations from the diagonalization of
the Hamiltonian (1) for $\varphi = \frac \pi 2$ in the case of
single crystal (solid lines), and for $\varphi =$ 0 and $\varphi =
\frac \pi 2$ for the oriented powder (dashed lines).}
\end{figure}

The position of the maximum of the resonant peaks (all of them
have been collected in Figure 4) has been fitted by using the
level structure (figure 2) resulted from the diagonalisation of
the Hamiltonian (1). In this fitting procedure we have taken into
account a small deviation, angle $\theta$, of the applied magnetic
field from the hard plane of the crystal. The angle $\theta$ of
misalignment was the only fitting parameter used in our
calculations. The resonant peaks for the oriented powder have been
fitted without considering such misalignment because in this case
there are always molecules, a fraction of $10^{-6}$ of the total
number of molecules in the sample, with their easy axis
perpendicular to the applied magnetic field. For both samples, we
have considered the occurrence of all possible transitions between
spin levels $i$ and $j$ such that $j-i=$ 1 and $j-i=$ 2. Each
$S_z$ level splits in two levels which, for example, at low fields
are the symmetric and antisymmetric combinations of $S_z$. The
levels appearing from $S_z$ = 10 are labeled 1 and 2, those from
$S_z$ = 9 are labeled 3 and 4 and so on. The results of our
fitting procedure are shown in Figure 4, solid lines correspond to
the single crystal and the dashed lines correspond to the powder.

Let us discuss first the data for the powder sample. In Figure 4
we show the fitting (dashed lines) for the resonant peaks at four
different frequencies (experimental values are open squares).
First, we observe the existence of two resonant field values for
each frequency which correspond to the energy absorption for the
cases when the magnetic field is parallel and perpendicular to the
hard axis, see Figure 1. The two peaks at 40 GHz observed at
$H_1=$ 3.2 T and $H_2=$ 4.9 T correspond to the quantum splitting
of $S_z = \pm$10 states for the two field orientations above
mentioned. Although we do not know the decoherence characteristic
time, the large values of the used frequency are likely to be in
the coherence range. Then the two resonant peaks correspond to the
quantum coherent oscillations of the spin between the levels
$S_z=$ +10 and $S_z=$ -10 in all molecules. At higher frequencies,
the resonant fields are larger than the barrier height and the
$ij$ transitions are the transitions between mixed levels, that
is, the levels on the right of line 2 in Figure 2.

\begin{figure}
\centerline{\epsfig{file=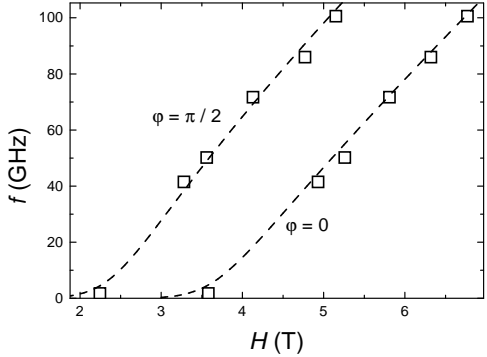,width=8truecm}}
\caption{\small Evolution of the quantum splitting of the ground
state as a function of the intensity of the tranverse field. Open
squares represent the experimental points for the two orientations
$\varphi =$ 0 and $\varphi = \frac \pi 2$ (powder sample) and the
dashed line is the theoretical fitting. The low frequency points
correspond to the frequency $f$ = 680 MHz and were extracted from
Ref. 14}
\end{figure}

We have also been able to fit all the resonances observed in the
single crystal for 40 different frequencies. In this case, we also
have the angle $\theta$ of misalignment, which is also the only
fitting parameter. The angle $\varphi$ takes only one value,
$\varphi$ = $\frac \pi 2$, which is the angle between the applied
magnetic field and the $x$ axis of all molecules ($x$ axis of the
single crystal). The data are shown in Figure 4 in which the black
points are the experimental data and the solid lines are the
result of our fitting procedure after the diagonalisation of the
spin Hamiltonian for the field values and frequencies used. As it
is shown in that figure, all the transitions correspond to either
$ij$ transitions with $j$ - $i$ = 1 or $j$ - $i$ = 2. It is
important to notice that the transitions ij = 12, 34, 56, 78,
etc., observed for field values lower than the barrier height of
each level, correspond to the quantum splitting of different $S_z$
levels. We are also observing high frequency transitions at low
fields which correspond to transitions between different,
non-consecutive ($j$ - $i$ = 2), $S_z$ levels. Though the field
dependence of such transitions is complicate, our fitting
procedure was able to reproduce all the observed transitions.

In a previous paper \cite{delBarco} we discussed the results
obtained at much lower frequency, 680 MHz. In Figure 5 we show, in
a logarithmic plot, the resonant field values assigned to the
quantum splitting, $\Delta_{10}$, of the $S_z = \pm$10 states for
the two set of experiments. In the same Figure 5 we also show the
result for the fitting of these data by the theoretical dependence
of $\Delta_{10}$ on the transverse field. The agreement between
theory and experiment is remarkable, and confirms our
interpretation. Moreover, the resonant field values for the peaks
observed at 680 MHz are such that it is possible to characterize
the two levels between which the resonance is observed, as the
symmetric and antisymmetric combinations of classical spin states.

To conclude, we have observed resonant peaks in $Fe_8$ molecules
as a function of the magnetic field perpendicular to the easy
axes. Our experiments cover the range of frequencies between 37
GHz and 110 GHz which allowed us to detect coherent spin quantum
oscillations for different orientations of spin $S$ = 10.

E. del B. acknowledges support from the University of Barcelona.
J. T., and J. M. H. acknowledge support from the CICYT project No.
IN96-0027 and the CIRIT project No. 1996-PIRB-00050. This work was
supported in part by NHMFL/IHRP 500/5031, and one of us (EdB)
acknowledges some support from the NHMFL Visitors Program during
his stay. The NHMFL is supported by a contractual agreement
between the State of Florida and the National Science Foundation
under contract NSF-DMR-95-27035.

\end{document}